\documentclass[11pt,a4paper]{article}

\usepackage[truedimen,margin=34mm]{geometry} 

\usepackage{mathrsfs}
\usepackage{amssymb}
\usepackage{amsmath}
\usepackage{ascmac}
\usepackage{amsthm}
\usepackage[dvipdfmx]{graphicx}
\graphicspath{{./figures/}}
\usepackage{booktabs}
\usepackage{setspace}
\usepackage{color}

\usepackage{titlesec}
\titleformat*{\section}{\large\bfseries}
\titleformat*{\subsection}{\it}

\def\th{{\theta}}

\def\bphi{{\text{\boldmath $\phi$}}}
\def\bbe{{\text{\boldmath $\beta$}}}

\def\bphi{{\text{\boldmath $\phi$}}}

\def\thh{{\widehat \th}}

\def\bphih{{\widehat \bphi}}

\def\bphih{{\widehat \bphi}}

\def\tht{\widetilde{\th}}

\def\Ga{\Gamma}

\def\0{{\text{\boldmath $0$}}}
\def\1{{\text{\boldmath $1$}}}

\def\x{{\text{\boldmath $x$}}}

\def\Var{{\rm Var}}

\def\E{{\rm E}}

\def\Boot{\text{Boot}}

\def\mit{\vspace{-0.23cm}\item}

\title{{\bf On Bootstrap Averaging Empirical Bayes Estimators}\footnote{This version: \today}}

\date{}

\begin{document}

\maketitle

\vspace{-1cm}\noindent
SHONOSUKE SUGASAWA\\
{\it Risk Analysis Research Center, The Institute of Statistical Mathematics}\\

\vspace{0.4cm}\noindent
{\large\bf Abstract.}\ \ 
Parametric empirical Bayes (EB) estimators have been widely used in variety of fields including small area estimation, disease mapping.
Since EB estimator is constructed by plugging in the estimator of parameters in prior distributions, it might perform poorly if the estimator of parameters is unstable.
This can happen when the number of samples are small or moderate.
This paper suggests bootstrapping averaging approach, known as ``bagging" in machine learning literatures,  to improve the performances of EB estimators.
We consider two typical hierarchical models, two-stage normal hierarchical model and Poisson-gamma model, and compare the proposed method with the classical parametric EB method through simulation and empirical studies.


\section{Introduction}
The parametric empirical Bayes estimators (Morris, 1983) are known to be a useful method producing reliable estimates of multidimensional parameters.
This technique is widely used in variety of fields such as small area estimation (Rao and Molina, 2015) and disease mapping (Lawson, 2013).
Let $\th_1,\ldots,\th_m$ be the multiple parameters of interest, and $y_1,\ldots,y_m$ be the independent observations generated from the distribution $f_i(y_i|\th_i), \ i=1,\ldots,m$.
To carry out an empirical Bayes estimation, it is assumed that the parameters $\th_1,\ldots,\th_m$ independently follows the distribution $g(\th_i; \bphi)$, where $\bphi$ is a vector of unknown parameters.
Therefore, we obtain a two stage model:
\begin{equation}\label{model}
y_i|\th_i \sim f_i(y_i|\th_i), \ \ \ \ \ \th_i\sim g(\th_i;\bphi), \ \ \ \ i=1,\ldots,m,
\end{equation}
which are independent for $i=1,\ldots,m$.
Under the setting, the posterior distribution of $\th_i$ is given by 
$$
\pi(\theta_i |y_i;\bphi)=\frac{f_i(y_i | \theta_i)g(\theta_i;\bphi)}{\int f_i(y_i | \theta_i)g(\theta_i;\bphi)d\th_i}, \ \ \ \ i=1,\ldots,m.
$$
The Bayes estimator $\tht_i$ of $\th_i$ under squared error loss is the conditional expectation (posterior mean) of $\th_i$ given $y_i$, that is
\begin{equation}\label{BE}
\tht_i\equiv \E[\th_i|y_i;\bphi]=\frac{\int \th_if_i(y_i | \theta_i)g(\theta_i;\bphi)d\th_i}{\int f_i(y_i | \theta_i)g(\theta_i;\bphi)d\th_i}, \ \ \ \ i=1,\ldots,m.
\end{equation}
However, the Bayes estimator $\tht_i$ depends on unknown model parameters $\bphi$, which can be estimated from the marginal distribution of all the data $y=\{y_1,\ldots,y_m\}$, given by
$$
L(\bphi)=\prod_{i=1}^m\int f_i(y_i | \theta_i)g(\theta_i;\bphi)d\th_i.
$$
Using the marginal distribution of $y$, one can immediately define the maximum likelihood (ML) estimator as the maximizer of $L(\bphi)$.
Based on the estimator $\bphih$, we obtain the empirical Bayes (EB) estimator of $\th_i$ as $\thh_i=\E[\th_i|y_i;\bphih]$.

The variability of the EB estimator $\thh_i$ can be measured by the integrated mean squared error (MSE) $\E[(\thh_i-\th_i)^2]$, where the expectation is taken with respect to $\th_i$'s and $y_i$'s following the model (\ref{model}). 
Since $\tht_i$ is the conditional expectation as given in (\ref{BE}), the MSE can be decomposed as $\E[(\thh_i-\th_i)^2]=R_1+R_2$ with $R_1=\E[(\tht_i-\th_i)^2]$ and $R_2=\E[(\thh_i-\tht_i)^2]$.
The first term $R_1$ is not affected by the estimation of $\bphi$ whereas the second term $R_2$ reflects the variability of the ML estimator $\bphih$, so that the second term can be negligibly small when $m$ is large.
However, in many applications, $m$ might be small or moderate, in which the contribution of the second term to the MSE cannot be ignored.
Hence, the EB estimator might perform poorly depending on the ML estimator $\bphih$.
To overcome this problem, we propose to use the bootstrap averaging technique, known as ``bagging" (Breiman, 1996) in machine learning literatures.
This method produces many estimators based on bootstrap samples, and average them to produce a stable estimator.
We adapt the bagging method to the EB estimation to improve the performances of EB estimators under small or moderate $m$.

This paper is organized as follows:
In Section \ref{sec:bagging}, we consider mean squared errors of EB estimators and propose a bootstrap averaging empirical Bayes (BEB) estimator for decreasing the mean squared error. 
In Section \ref{sec:FH} and Section \ref{sec:PG}, we apply the BEB estimators in well-known two-stage normal hierarchical model and Poisson-gamma model, respectively, and compare the performances between BEB and EB estimators through simulation and empirical studies.
In Section \ref{sec:conc}, we provide conclusions and discussions.

\section{Bootstrap Averaging Empirical Bayes Estimators}\label{sec:bagging}

As noted in the previous section, the performances of the EB estimators depend on the variability of the estimator $\bphih$, which cannot be ignored when $m$ is not large.
To reduce the variability of the empirical Bayes estimator $\thh_i$, we propose to average many empirical Bayes estimators with bootstrap estimates of $\bphi$ rather than computing one empirical Bayes estimator from the observation $Y=\{y_1,\ldots,y_m\}$.
Specifically, letting $Y_{(b)}=\{y_1^{(b)},\ldots,y_m^{(b)}\}$ be a bootstrap samples of the original observation $Y$, we define $\bphih_{(b)}$ be an estimator of $\bphi$ based on the bootstrap sample $Y_{(b)}$.
Then the bagging empirical Bayes (BEB) estimator is given by
\begin{equation}\label{BEB}
\thh_i^{\Boot}=\frac1B\sum_{b=1}^B\tht_i(y_i,\bphih_{(b)}).
\end{equation}

Similarly to Breiman (1996), we note that 
\begin{align*}
\frac1B\sum_{b=1}^B\left\{\tht_i(y_i,\bphih_{(b)})-\th_i\right\}^2
&=\frac1B\sum_{b=1}^B\tht_i(y_i,\bphih_{(b)})^2-2\thh_i^{\Boot}\th_i+\th_i^2\\
&\geq \bigg\{\frac1B\sum_{b=1}^B\tht_i(y_i,\bphih_{(b)})\bigg\}^2-2\thh_i^{\Boot}\th_i+\th_i^2
=(\thh_i^{\Boot}-\th_i)^2.
\end{align*}
By taking expectation with respect to the model (\ref{model}), we have
$$
\frac1B\sum_{b=1}^B\E\left[\left\{\tht_i(y_i,\bphih_{(b)})-\th_i\right\}^2\right]
\geq 
\E\left[(\thh_i^{\Boot}-\th_i)^2\right],
$$
which means that the integrated MSE of BEB estimator (\ref{BEB}) is smaller than bootstrap average of the integrated MSE of the EB estimator.
Hence, the BEB estimator is expected to perform better than the EB estimator.
The amount of improvement depends on   
$$
\frac1B\sum_{b=1}^B\tht_i(y_i,\bphih_{(b)})^2-\bigg\{\frac1B\sum_{b=1}^B\tht_i(y_i,\bphih_{(b)})\bigg\}^2
=\frac1B\sum_{b=1}^B\left\{\tht_i(y_i,\bphih_{(b)})-\thh_i^{\Boot}\right\}^2,
$$
which is the bootstrap variance of the EB estimator and it vanishes as $m\to\infty$ but it would not be negligible when $m$ is not large.
Therefore, when $m$ is small or moderate, the BEB estimator would improve the performance of the EB estimator.
In the subsequent section, we investigate the performances of the EBE estimator compared with the EB estimator in the widely-used hierarchical models.

\section{Two-stage normal hierarchical model}\label{sec:FH}


\subsection{Model description}
We first consider the two-stage normal hierarchal model to demonstrate the proposed bagging procedure.
The two-stage normal hierarchical model is described as 
\begin{equation}\label{FH}
y_i|\th_i\sim N(\th_i, D_i), \ \ \ \ \ \ 
\th_i\sim N(\x_i^t\bbe,A), \ \ \ \ i=1,\ldots,m,
\end{equation}
where $D_i$ is known sampling variance, $\x_i$ and $\bbe$ are a vector of covariates and regression coefficients, respectively, $A$ is an unknown variance. 
Let $\bphi=(\bbe^t,A)^t$ be the vector of unknown parameters.
The model (\ref{FH}) is known as the Fay-Herriot model (Fay and Herriot, 1979) in the context of small area estimation.

Under the model (\ref{FH}), the Bayes estimator of $\th_i$ is
\begin{equation*}
\tht_i(y_i;\bphi)=\x_i^t\bbe+\frac{D_i}{A+D_i}(y_i-\x_i^t\bbe).
\end{equation*}
Concerning the estimation of unknown parameter $\bphi$, we here consider the maximum likelihood estimator for simplicity.
Since $y_i\sim N(\x_i^t\bbe,A+D_i)$ under the model (\ref{FH}), the maximum likelihood estimator $\bphih$ is defined as the maximizer of the function:
\begin{equation*}
Q(\bphi)=\sum_{i=1}^m\log(A+D_i)+\sum_{i=1}^m\frac{(y_i-\x_i^t\bbe)^2}{A+D_i}.
\end{equation*}
While several other estimating methods are available, we here only consider the maximum likelihood estimator for presentational simplicity.
Using the maximum likelihood estimator $\bphih$, we obtain the EB estimator of $\th_i$ as $\tht_i(y_i;\bphih)$.


\subsection{Simulation study}\label{sec:FHsim}
We here evaluate the performances of the BEB estimator together with the EB estimator under the normal hierarchical model (\ref{FH}) without covariates, namely $\x_i^t\bbe=\mu$.
We considered $m=10, 15,\ldots,40$.
For each $m$, we set $D_i$ as equally spaced points between $0.5$ and $1.5$.
Concerning the true parameter values, we used $\mu=0$ and four cases for $A$, namely $A=0.1, 0.3, 0.5$ and $0.7$.
The simulated data was generated from the model (\ref{FH}) in each iteration, and computed the EB and BEB estimates of $\th_i$.
Based on $R=5000$ simulation runs we calculated the simulated mean squared errors (MSE) defined as 
\begin{equation}\label{sim-mse}
\text{MSE}=\frac{1}{mR}\sum_{i=1}^m\sum_{r=1}^R(\thh_i^{(r)}-\th_i^{(r)})^2,
\end{equation} 
where $\thh_i^{(r)}$ is the EBE or EB estimates and $\th_i^{(r)}$ is the true value of $\th_i$ in the $r$th iteration.

\begin{figure}[!htb]
\hspace{-0.5cm}
\includegraphics[width=15cm]{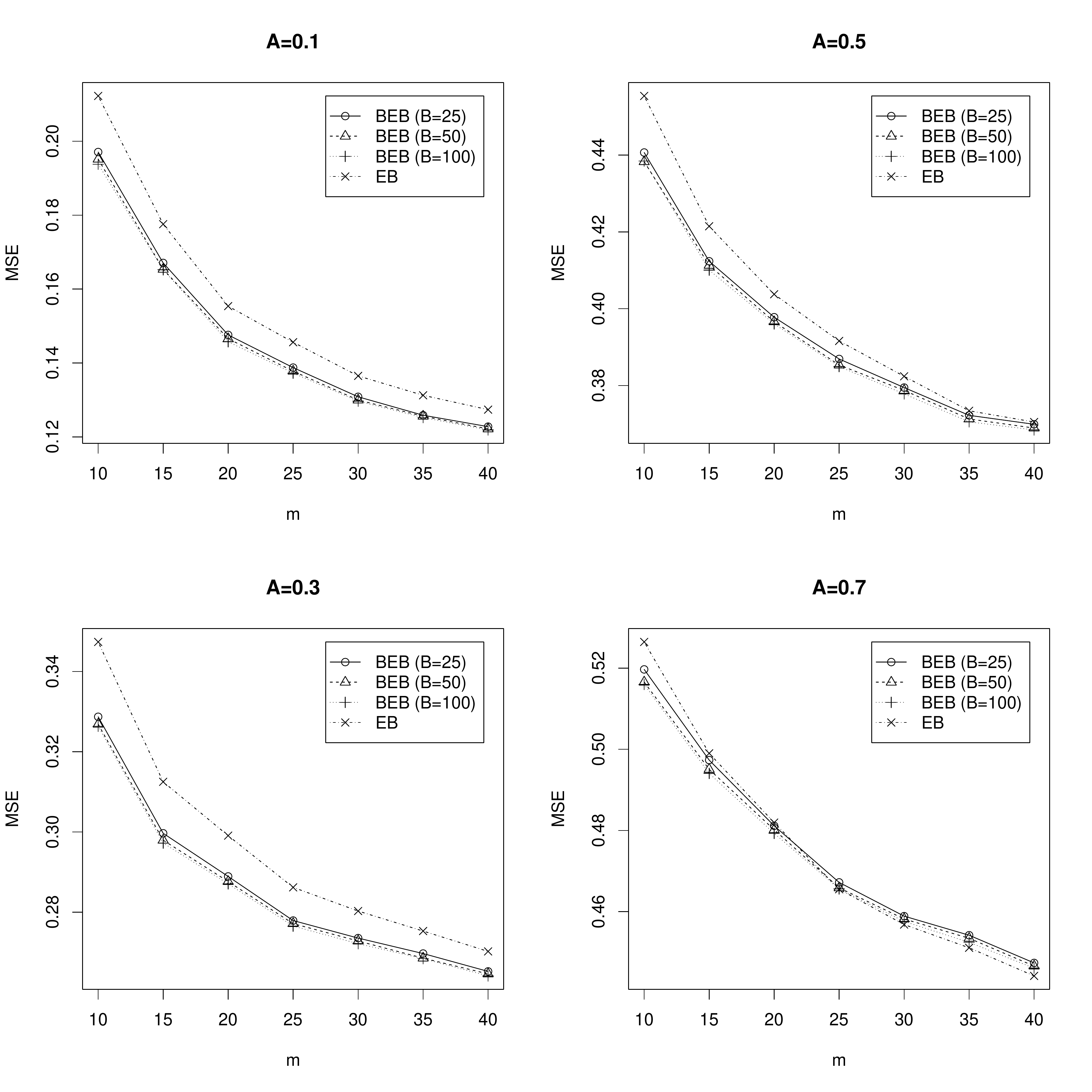}
\caption{The simulated MSE of three estimators, BEB (bootstrap averaging empirical Bayes estimator) and EB (empirical Bayes estimator) in two-stage normal hierarchical model.}
\label{fig:FH}
\end{figure}

In Figure \ref{fig:FH}, we present the simulated MSE of the EB estimator as well as the three BEB estimator using $25, 50$ and $100$ bootstrap samples under various settings of $A$ and $m$.
It is observed that the BEB estimator performs better than the EB estimator on the whole.  
In particular, the improvement is greater when $A$ is small compared with $D_i$, which is often arisen in practice.
Moreover, as the number of $m$ gets larger, the MSE differences get smaller since the variability of estimating $\bphi$ vanishes when $m$ is sufficiently large.
We also found that the ML estimator of $A$ often produces $0$ estimates when $m$ is small, in which the EB estimator is known to perform poorly.
However, the BEB estimator can avoid the problem since the BEB estimator is aggregated by $B$ bootstrap estimators and at least one bootstrap estimates should be non-zero.
In fact, by investigating the case where the ML estimator produces $0$ estimates of $A$, the some bootstrap estimates of $A$ were away from $0$.
This would be one of the reason why the BEB estimator performs better than the EB estimator in this setting.


\subsection{Example: corn data}\label{sec:corn}
\begin{figure}[!htb]
\hspace{-0.5cm}
\includegraphics[width=15cm]{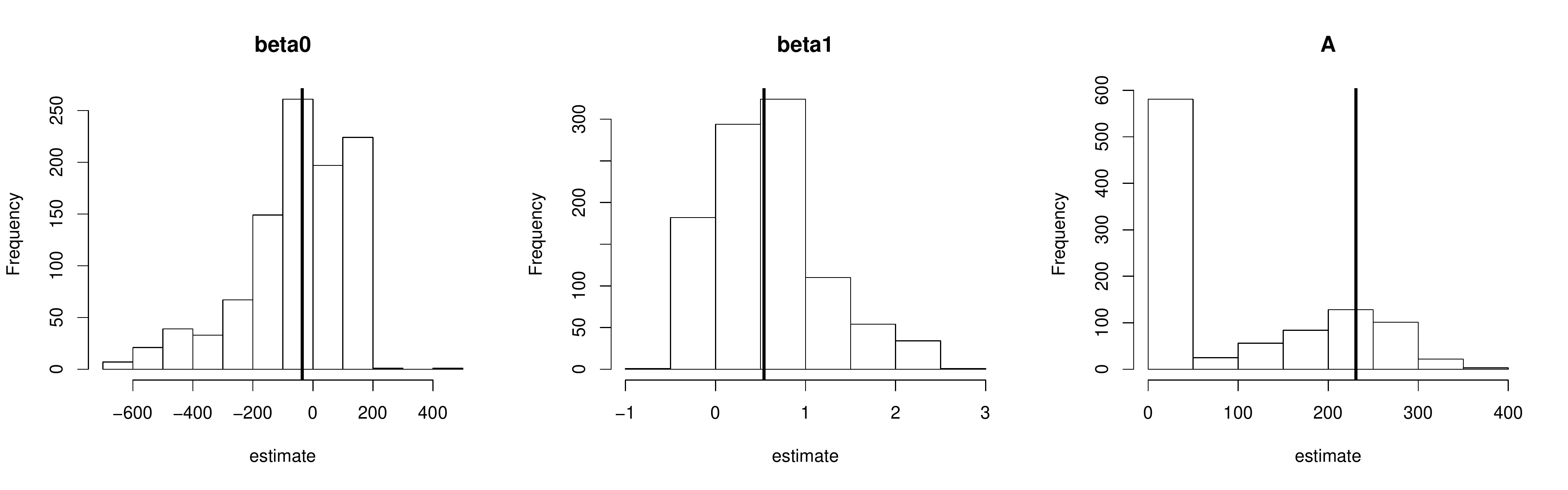}
\caption{The histograms of the bootstrap estimates of $\beta_0$ (left), $\beta_1$ (center) and $A$ (right).
Each vertical line denotes the maximum likelihood estimate.}
\label{fig:Corn-para}
\end{figure}

We next illustrate the performances of the BEB estimator by using the corn and soybean productions in 12 Iowa counties, which has been used as an example in the context of small area estimation.
Especially, we use the area-level data set given in table 6 in Dass et al. (2012) and we here focus only on corn productions for simplicity.
The data set consists of $m=8$ areas with sample sizes in each area ranging from 3 to 5, and survey data of corn production $y_i$, sampling variance $D_i$ and the satellite data of corn $x_i$ as the covariate observed in each area.
We considered the following hierarchical model:
\begin{equation}\label{FH}
y_i|\th_i\sim N(\th_i,D_i), \ \ \ \ \ \th_i\sim N(\beta_0+\beta_1x_i,A), \ \ \ \ i=1,\ldots,m,
\end{equation}
where $\beta_0,\beta_1$ and $A$ are unknown parameters.
For the data set, we computed the BEB as well as EB estimators.
We used $1000$ bootstrap samples for computing the BEB estimator.
In Figure \ref{fig:Corn-para}, we present the histogram of of the bootstrap estimates used in the BEB estimates and the maximum likelihood (ML) estimates used in the EB estimators.
We can observe that the bootstrap estimates vary depending on the bootstrap samples.
Moreover, in Table \ref{tab:corn-est}, we show the BEB and EB estimates of $\th_i$, which shows that the BEB estimator produces different estimates from the EB estimator since the number of areas $m$ is only $8$.

\begin{table}[!htb]
\centering
\caption{Direct estimates (DE) $y_i$, standard deviation (SV) $\sqrt{D_i}$, empirical Bayes (EB) estimates and bagging empirical Bayes (BEB) estimates in each county.
\label{tab:corn-est}
}
\medskip
\begin{tabular}{ccccccc}
  \hline
 & DE & SD & EB & BEB \\ 
  \hline
Franklin & 158.62 & 5.70 & 155.79 & 141.08 \\ 
  Pocahontas & 102.52 & 43.41 & 102.82 & 97.48 \\ 
  Winnebago & 112.77 & 30.55 & 119.74 & 117.34 \\ 
  Wright & 144.30 & 54.00 & 127.86 & 124.05 \\ 
  Webster & 117.59 & 21.30 & 109.61 & 102.42 \\ 
  Hancock & 109.38 & 15.66 & 121.84 & 126.51 \\ 
  Kossuth & 110.25 & 12.11 & 116.05 & 118.53 \\ 
  Hardin & 120.05 & 36.81 & 136.97 & 137.05 \\ 
   \hline
\end{tabular}
\end{table}

\section{Poisson-gamma model}\label{sec:PG}


\subsection{Setup}
The Poisson-gamma model (Clayton and Kalder, 1987) is described as
\begin{equation}\label{PG}
z_i|\th_i\sim \text{Po}(n_i\th_i), \ \ \ \ \ 
\th_i\sim \Gamma(\nu m_i,\nu), \ \ \ \ i=1,\ldots,m,
\end{equation}
where $m_i=\exp(\x_i^t\bbe)$, $\x_i$ and $\bbe$ are a vector of covariates and regression coefficients, respectively, $\nu$ is an unknown scale parameter. 
This model is used as the standard method of disease mapping.
Let $\bphi=(\bbe^t,\nu)^t$ be the vector of unknown parameters.
The model (\ref{PG}) is known as the Poisson-Gamma model considered in Clayton and Kaldor (1987) and used in disease mapping.

Under the model (\ref{PG}), the Bayes estimator of $\th_i$ is given by
$$
\tht_i(y_i;\bphi)=\frac{z_i+\nu m_i}{n_i+\nu}.
$$
Since the Bayes estimator depends on unknown $\bphi$, we need to replace $\bphi$ by its estimator.
Noting that the gamma prior of $\th_i$ is a conjugate prior for the mean parameter in the Poisson distribution, the marginal distribution of $y_i$ is the negative binomial distribution with the probability function:
$$
f_m(y_i;\bphi)=\frac{\Gamma(z_i+\nu m_i)}{\Gamma(z_i+1)\Gamma(\nu m_i)}\left(\frac{n_i}{n_i+\nu}\right)^{z_i}\left(\frac{\nu}{n_i+\nu}\right)^{\nu m_i}.
$$
Then the maximum likelihood estimator of $\bphi$ is defined as $\bphih=\text{argmax}_{\bphi} \sum_{i=1}^m \log f_m(y_i;\bphi)$, which enables us to obtain the empirical Bayes estimator $\tht_i(y_i;\bphih)$.


\subsection{Simulation study}
\begin{figure}[!htb]
\hspace{-0.5cm}
\includegraphics[width=15cm]{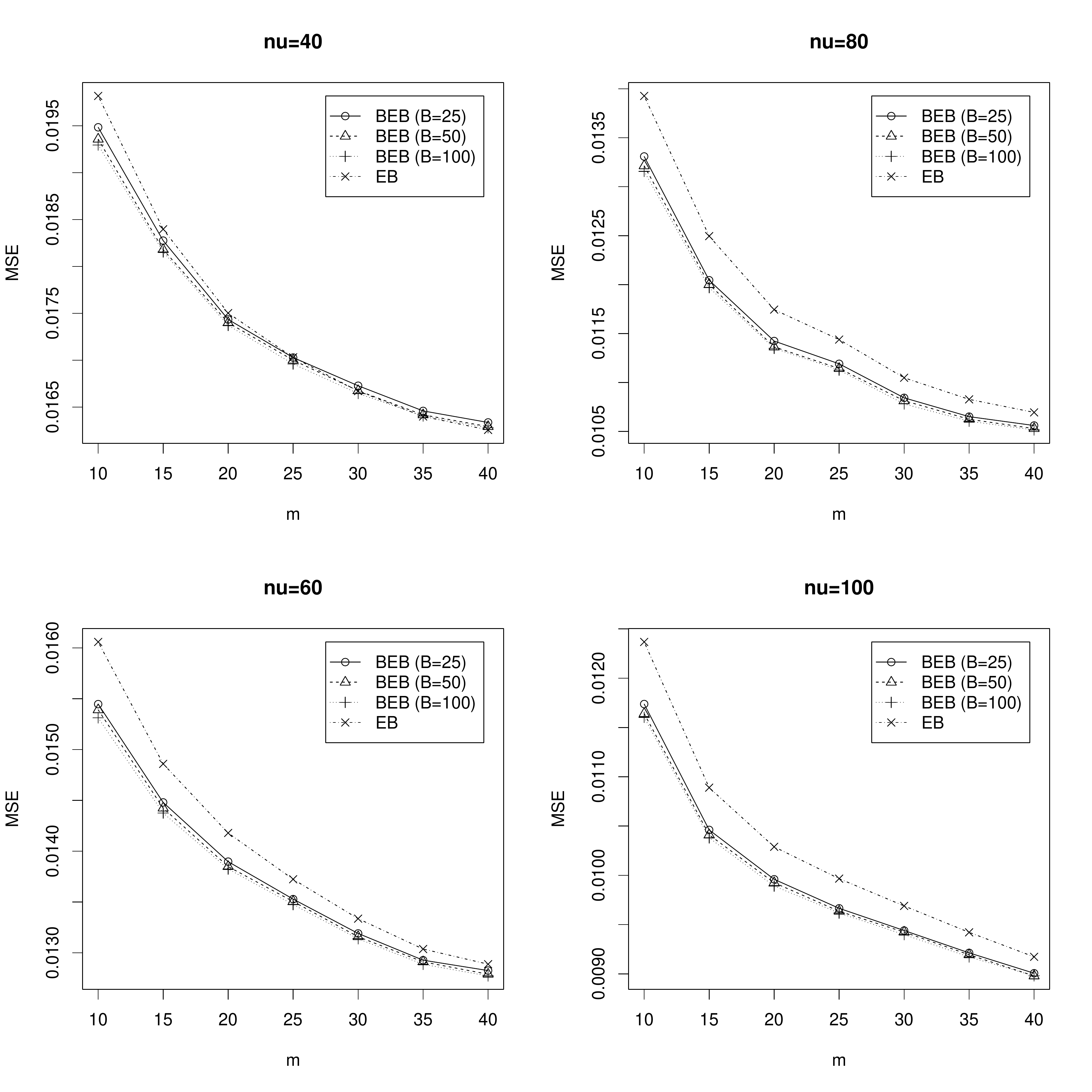}
\caption{The simulated MSE of three estimators, BEB (bootstrap averaging empirical Bayes estimator) and EB (empirical Bayes estimator) in Poisson-gamma model.}
\label{fig:PG}
\end{figure}

We next evaluated the performances of the BEB estimator under the Poisson-gamma model without covariates, described as 
\begin{equation}\label{PG-sim}
z_i|\th_i\sim \text{Po}(n_i\th_i), \ \ \ \th_i\sim \Ga(\nu\mu,\nu), \ \ \ \ \  i=1,\ldots,m,
\end{equation}
where we set $\mu=1$ and $\nu=40, 60, 80$ and $100$.
Note that $\nu$ is a scale parameter and $\Var(\th_i)=\mu/\nu$, so that random effect variance $\Var(\th_i)$ is a decreasing function of $\nu$.
Regarding the number of areas, we considered $m=10,15,\ldots,40$.
For each $m$, we set $n_i$ as rounded integers of equally spaced numbers between $10$ and $50$. 
Similarly to Section \ref{sec:FHsim}, using (\ref{sim-mse}) with $R=5000$ simulation runs, we calculated the MSE of the BEB estimator as well as the EB estimator of $\th_i$. 
The results are presented in Figure \ref{fig:PG}, which show that the BEB estimator tends to perform better than the EB estimator.
In particular, the amount of improvement is greater when $m$ is not large as we expected.
Moreover, we can also observe that the MSE difference tends larger as $\nu$ gets larger, which corresponds to the case where the random effect variance gets smaller.
This is consistent to the results in the normal model given in Section \ref{sec:FHsim}.

\subsection{Example: Scottish lip cancer}
\begin{figure}[!htb]
\hspace{-0.5cm}
\includegraphics[width=15cm]{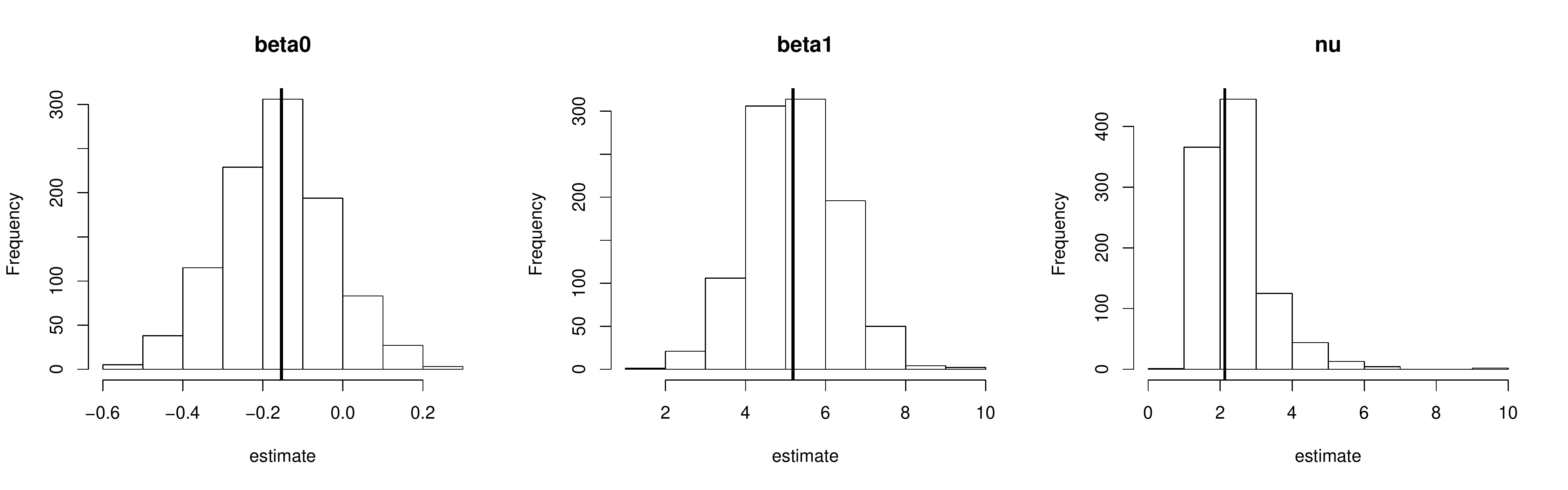}
\caption{The histograms of the bootstrap estimates of $\beta_0$ (left), $\beta_1$ (center) and $\nu$ (right).
Each vertical line denotes the maximum likelihood estimate. }
\label{fig:Scott-para}
\end{figure}

\begin{figure}[!htb]
\centering
\includegraphics[width=10cm]{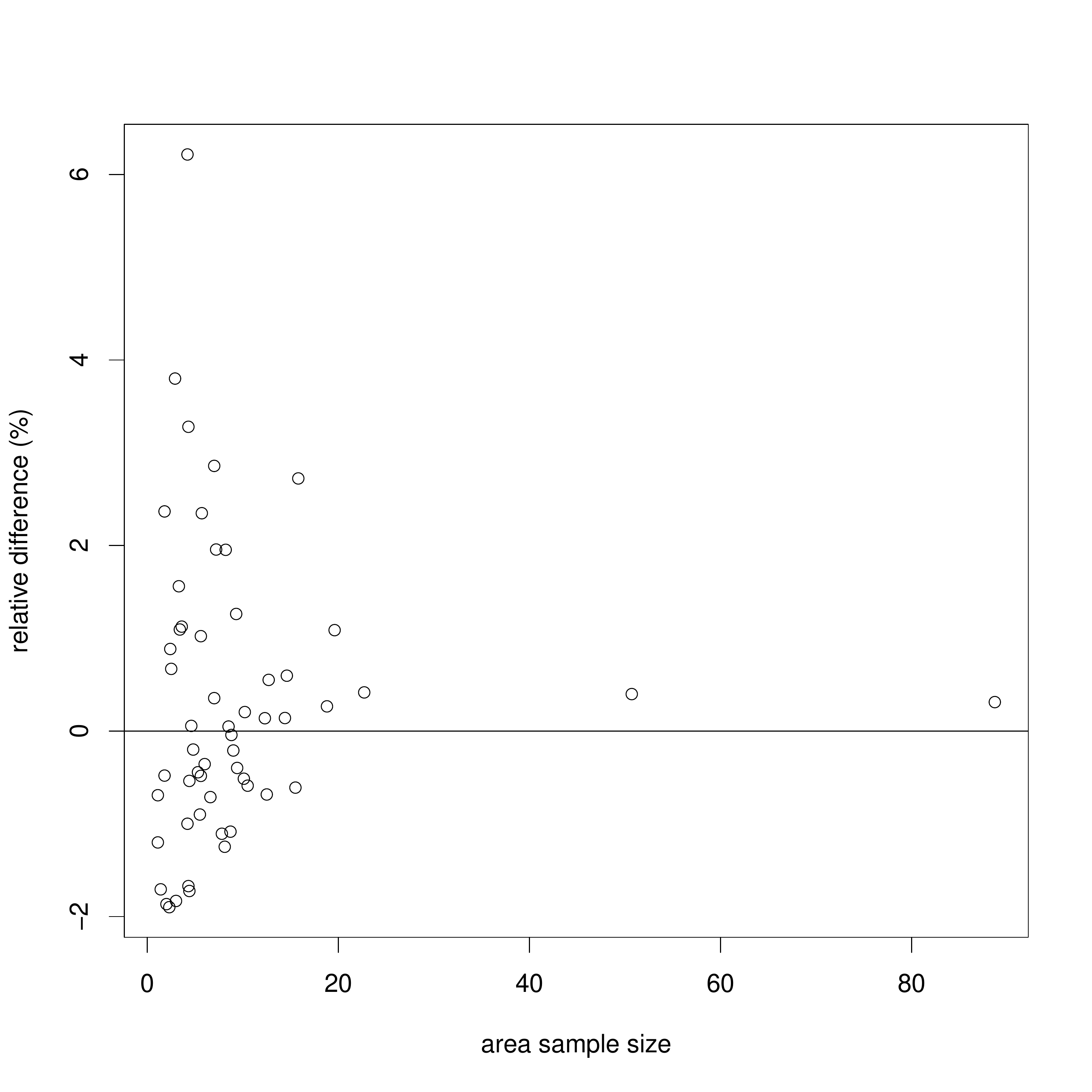}
\caption{The scatter plot of the area sample size $n_i$ against the percent relative difference between the BEB and EB estimates of $\th_i$.}
\label{fig:Scott-est}
\end{figure}

We applied the BEB and EB method to the famous Scottish lip cancer data during the 6 years from 1975 to 1980 in each of the $m=56$ counties of Scotland.
For each county, the observed and expected number of cases are available, which are respectively denoted by $z_i$ and $n_i$.
Moreover, the proportion of the population employed in agriculture, fishing, or forestry is available for each county, thereby we used it as a covariate $\text{AFF}_i$, following Wakefield (2007). 
For each area, $i=1,\ldots,m$, we consider the Poisson-gamma model:
\begin{equation}\label{PG}
z_i|\th_i\sim \text{Po}(n_i\th_i), \ \ \ \th_i\sim \Gamma(\nu\exp(\beta_0+\beta_1\text{AFF}_i),\nu),
\end{equation}
where $\th_i$ is the true risk of lip cancer in the $i$th area, and $\beta_0,\beta_1$ and $A$ are unknown parameters.
For the data set, we computed the BEB as well as EB estimates of $\th_i$, where we used $1000$ bootstrap samples for computing the BEB estimator.
In Figure \ref{fig:Scott-para}, we present the quantiles of the bootstrap estimates used in the BEB estimates and the maximum likelihood (ML) estimates used in the EB estimators.
We can observe that the bootstrap estimates vary depending on the bootstrap samples while the variability seems small compared with Figure \ref{fig:Corn-para}.
This might comes from that the number of areas in this case is much larger than the corn data in Section \ref{sec:corn}.
Finally, in Figure \ref{fig:Scott-est}, we show the scatter plot of percent relative difference between the BEB and EB estimates, that is, $100(\thh^{\Boot}-\thh_i)/\thh_i$, against the number of expected number of cases $n_i$.
Figure \ref{fig:Scott-est} shows that the differences get larger as $n_i$ gets small since the direct estimator $y_i=z_i/n_i$ of $\th_i$ is shrunk toward the regression mean $\exp(\beta_0+\beta_1\text{AFF}_i)$ in areas with small $n_i$.

\section{Conclusion and Discussion}\label{sec:conc}
We have proposed the use of bootstrap averaging, known as ``bagging" in the context of machine learning, for improving the performances of empirical Bayes (EB) estimators.
We focused on two models extensively used in practice, two-stage normal hierarchical model and Poisson-gamma model.
In both models, the simulation studies revealed that the bootstrap averaging EB (BEB) estimator performs better than the EB estimator.

In this paper, we considered the typical area-level models as an application of the BEB estimator.
However, the BEB method would be extended to the more general case, for example, generalized linear mixed models.
The detailed comparison in such models will be left to a future study.

\medskip

\end{document}